\documentclass[showpacs,amsmath,amssymb,aps,showkeys,floatfix,a4paper,nofootinbib]{revtex4}

\usepackage{dcolumn}
\usepackage{bm}
\usepackage{epsfig}
\usepackage{amsfonts}
\usepackage{amssymb,amscd}
\usepackage{graphicx}
\usepackage{wrapfig}

\usepackage{comment}
\usepackage{color}

\def\lsim{\raise0.3ex\hbox{$<$\kern-0.75em\raise-1.1ex\hbox{$\sim$}}}
\def\gsim{\raise0.3ex\hbox{$>$\kern-0.75em\raise-1.1ex\hbox{$\sim$}}}

\def\beq{\begin{equation}}
\def\eeq{\end{equation}}
\def\bea{\begin{eqnarray}}
\def\eea{\end{eqnarray}}
\def\bq{\begin{quote}}
\def\eq{\end{quote}}

\newcommand{\sigmatot}{\sigma_\text{tot}}
\newcommand{\sigmael}{\sigma_\text{el}}
\newcommand{\sigmadiff}{\sigma_\text{diff}}

\DeclareMathOperator{\Real}{\text{Re}}

\def\gappeq{\mathrel{\rlap {\raise.5ex\hbox{$>$}}
{\lower.5ex\hbox{$\sim$}}}}

\def\lappeq{\mathrel{\rlap{\raise.5ex\hbox{$<$}}
{\lower.5ex\hbox{$\sim$}}}}

\def\Toprel#1\over#2{\mathrel{\mathop{#2}\limits^{#1}}}

\begin{document}


\title{Diffractive excitation in $pp$ and $pA$ collisions at high energies}
\author{V.~P. Gon\c{c}alves}
\email{barros@ufpel.edu.br}
\affiliation{High and Medium Energy Group, \\
Instituto de F\'{\i}sica e Matem\'atica, Universidade Federal de Pelotas\\
Caixa Postal 354, CEP 96010-900, Pelotas, RS, Brazil}
\author{R.~Palota da Silva}
\email{palota.rafael@gmail.com}
\affiliation{High and Medium Energy Group, \\
Instituto de F\'{\i}sica e Matem\'atica, Universidade Federal de Pelotas\\
Caixa Postal 354, CEP 96010-900, Pelotas, RS, Brazil}
\author{P. V. R. G. Silva}
\email{pvrgsilva@ufpel.edu.br}
\affiliation{High and Medium Energy Group, \\
Instituto de F\'{\i}sica e Matem\'atica, Universidade Federal de Pelotas\\
Caixa Postal 354, CEP 96010-900, Pelotas, RS, Brazil}
\date{\today}

\begin{abstract}
In this paper we consider the Good - Walker approach for the diffractive excitation and updated the Miettinen -- Pumplin (MP) model for $pp/\bar{p}p$ collisions considering the recent LHC data for the total and elastic $pp$ cross sections. 
 The behavior of the total, elastic and diffractive cross sections is analyzed and predictions for the energies of Run 3 of the LHC and those of the Cosmic Rays experiments are derived. Our results demonstrate that the MP model is able to describe the current data and that it implies that the cross section for the diffraction excitation in $pp$ collisions is almost constant in the energy range probed by the LHC and slowly decreases at higher energies. Our results indicate that the Pumplin bound is not reached at the LHC and Cosmic Ray energies. Moreover, the implications of the diffractive excitation in $pA$ collisions is discussed. In particular, the  MP model, constrained by the $pp$ data, is   used to derive the main quantities present in the treatment of the diffractive excitation in $pA$ collisions. Predictions for the total, elastic and diffractive $pA$ cross sections are presented considering different nuclei. We demonstrate that the effect of fluctuations decreases at larger energies and heavier nuclei. The energy dependence of the diffractive excitation cross section in $pA$ collisions is estimated for different nuclei and compared with the predictions for the proton dissociation induced by photon interactions.
 \end{abstract}

\pacs{12.38.Aw, 13.85.Lg, 13.85.Ni}
\keywords{Quantum Chromodynamics, Diffractive Excitation, Hadronic Collisions}

\maketitle

\section{Introduction}
\label{intro}
The recent LHC measurements for the total, elastic and differential cross sections at high energies of 2.76, 7, 8 and 13 TeV  
\cite{Antchev:2011,Antchev:2013a,Antchev:2013c,Antchev:2013b,Antchev:2015,Antchev:2016,Deile:2017,Antchev:2017b,Antchev:2017,Nemes_talk:2018,Antchev:2018a,Antchev:2018b,Aad:2014,Aaboud:2016} have motivated 
an intense debate about the treatment of the hadronic interactions at high energies (For a recent review see, e.g. Ref. \cite{Pancheri:2016yel}). As hadrons are composite objects, it is natural to expect that the description of these interactions could be performed  in terms of elementary interactions between their parton constituents. During the last decades, several models have been proposed, considering different assumptions and ingredients to treat the hadronic structure, the interactions between its constituents and the unitarization effects (See e.g. Refs. 
\cite{Good:1960,Chou:1968bc,Chou:1968bg,Bialas:1972cx,Bialas:1972ru,Fialkowski:1975ta,VanHove:1976fi,VanHove:1977kc,Miettinen:1978,Nussinov:1979ca,Miettinen:1979ns,Bertsch:1981py,Alberi:1981sz,DiasdeDeus:1987yw,Margolis:1988ws,Durand:1988ax,Heiselberg:1991is,Fletcher:1992sy,Blaettel:1993,Godbole:2004kx}). Some of them have been recently updated using the LHC data and some new approaches have been proposed (See e.g. Refs.    
\cite{Sapeta:2004av,Sapeta:2005,Guzey:2006,Kopeliovich:2005us,Avsar:2007xg,Ryskin:2007qx,Flensburg:2010kq,Lipari:2009rm,Fagundes:2015vba,Bierlich:2016smv,heikke,cepila}).  Although the total and elastic cross sections are dominated by nonperturbative contributions, the energy rising is expected to be driven by hard partonic subcollisions. Another important consequence of internal degrees of freedom is the diffractive excitation of the colliding particles, in which one (or both) particle(s) is (are) excited to a higher mass state with the same quantum numbers. In the case of single diffraction, the other initial state hadron remains intact. As demonstrated many years ago by Good and Walker~\cite{Good:1960}, the diffractive excitation arise from the fluctuating structure of the hadron, with its description being determined by the eigenstates of the scattering operator, which are the basis to express the physical states. Therefore, the diffractive excitation processes can be considered a direct probe of the proton wave function. Two other motivations to study the diffractive excitation processes are associated to the fact that its contribution for the hadronic cross sections is non-negligible and that its energy behaviour is expected to be strongly sensitive to the approach of the black disc limit, which reduces the fluctuations in the eigenstates (See e.g. Ref. \cite{Pancheri:2016yel}).

Our goal in this paper is to present predictions for the diffractive excitation in $pp$ and $pA$ collisions for the LHC and ultra high energy cosmic rays ranges. Motivated by the recent LHC data for the total and elastic proton -- proton cross sections at $\sqrt{s} =$ 8 and 13 TeV, we update previous studies \cite{Sapeta:2004av,Sapeta:2005} of the diffractive excitation $pp$ cross section ($\sigmadiff$) using the Miettinen -- Pumplin (MP) model \cite{Miettinen:1978}, which assumes that the scattering eigenstates correspond to parton showers, which interact via parton - parton scattering. The LHC data for $\sigmatot$ and $\sigmael$ are used to constrain the energy dependence of the main parameters of the MP model, and a parametrization for this dependence is derived. Parameter free predictions for $\sigmadiff$ are calculated and the results are compared with the current experimental data. We demonstrate that the MP model is able to describe the data and that the contribution of the diffractive excitation slowly decreases at higher energies. Moreover, the dependence on the impact parameter ($b$)  of the elastic and diffractive cross sections for different energies is estimated. In agreement with previous studies \cite{Miettinen:1978,Sapeta:2005}, our results demonstrate that the diffractive excitation is dominantly peripheral, with the maximum of the cross section ocurring at larger impact parameters when the energy is increased. Finally, using our results for $\sigmatot$, $\sigmael$ and $\sigmadiff$ we compare our predictions for the ratio $R(s) \equiv (\sigmael + \sigmadiff)/\sigmatot$ with the Pumplin bound \cite{Pumplin:1973}, which established that this ratio should be smaller than $1/2$. We demonstrate that $R \approx 0.35$ at LHC energies, with the increasing with the energy being very slow. 

For the case of $pA$ collisions, Gribov \cite{gribov} have demonstrated that the diffractive excitation gives a significant contribution for the treatment of the multiple scatterings proposed by Glauber \cite{Glauber:1970}. Different approaches to include these corrections in the Glauber formalism have been formulated during the last decades. In this paper, we will consider the model proposed in Refs. \cite{Heiselberg:1991is,Blaettel:1993} and discussed in detail in a series of papers (See e.g. Refs. \cite{Guzey:2006,Frankfurt:2011cs}). Such model is based on the Good -- Walker formalism  and takes into account of the cross section fluctuations. One of the main inputs in this model is the broadness of the cross section fluctuation around the average value, denoted $\omega_{\sigma}$ in Refs. \cite{Heiselberg:1991is,Blaettel:1993} and below, which can be determined by the forward diffractive and elastic $pp$ cross sections. From our analysis of the $pp$ scattering using the Miettinen -- Pumplin model, such quantity can be directly calculated and its energy dependence is unambiguously determined. As a consequence, parameter free predictions for the total, elastic and diffractive excitation cross sections can be obtained. We will present our predictions for the energy dependence of these cross sections considering proton -- lead ($pPb$) and proton -- nitrogen ($pN$) collisions. The impact of the fluctuations in the total and elastic cross sections is estimated, and a comparison with previous results is presented. Finally, in the case of the diffractive excitation, we also compare our results with the predictions for the proton dissociation by photon -- induced interactions.

The paper is organized as follows. In the next Section, we present a brief review of the Miettinen -- Pumplin model and determine its main parameters using the recent LHC data. A parametrization for the energy dependence of these parameters is proposed and the predictions with the experimental data for the diffractive excitation is presented. In Section \ref{sec:difpA}, we discuss the inclusion of the fluctuations in the $pA$ cross section, as proposed in Refs. \cite{Heiselberg:1991is,Blaettel:1993}, and the parameter $\omega_{\sigma}$ and its energy dependence is determined using the MP model. Predictions for the $pA$ cross sections are presented and the impact of the fluctuations is estimated. Finally, in Section \ref{sec:Sum}, our main conclusions are summarized.

\section{Diffractive excitation in $pp$ collisions}
\label{sec:difpp}

The model for diffractive excitation introduced by Miettinen and Pumplin (MP)~\cite{Miettinen:1978} is based on the Good-Walker approach~\cite{Good:1960}, which consists in express the physical state of the incoming (beam) hadron $|H\rangle$ as a superposition of eigenstates $\{ |t_k\rangle \}$ of the scattering operator $\hat{T}$, which form a complete set of normalized states, such that $|H \rangle = \sum_k C_k |t_k \rangle$ and $\mbox{Im} \hat{T} |t_k\rangle = t_k |t_k\rangle$ with $ 0 \le t_k \le 1$  due to the unitarity. These eigenstates can only go through elastic scattering and, moreover, they interact with different intensities (eingenvalues) with the target. It is precisely these differences in the intensities that originate the diffractive excitation present in the final state of composed particles scattering. The MP model considers the eigenstates to be related to the partons inside the hadrons,
and their degrees of freedom are described in terms of their impact parameter $\mathbf{b}_i$
inside the hadron and their rapidities $y_i$, i.e. $|t_k\rangle \equiv \left|\mathbf{b}_1,\dots,\mathbf{b}_N;y_1,\dots,y_N\right\rangle$. Each eingestates may contain $N$ partons and the 
state of the incoming hadron $|H\rangle$ is written as
\begin{equation}
 \left|H\right \rangle = \sum_N \prod_{i=1}^N \int d^2\mathbf{b}_idy_i C_N(\mathbf{b}_1,\dots,\mathbf{b}_N;y_1,\dots,y_N)\left|\mathbf{b}_1,\dots,\mathbf{b}_N;y_1,\dots,y_N\right\rangle.
\end{equation}
One of the basic assumptions in the MP model is that the partons in the hadrons are  uncorrelated, so that the probability to find
such state of $N$ partons in the wave function is given by a Poisson distribution
\begin{equation}
 |C_N(\mathbf{b}_1,\dots,\mathbf{b}_N;y_1,\dots,y_N)|^2 = e^{-G^2}\left(\frac{G^{2N}}{N!}\right)\prod_{i=1}^N|C_i(\mathbf{b}_i,y_i)|^2,
\end{equation}

\noindent where $G^2$ is the average number of partons in the
eigenstate. Moreover, the probability to find a parton
is assumed to follow a Gaussian distribution in rapidity and in impact parameter
\begin{equation}
 |C_i(\mathbf{b}_i,y_i)|^2 = \frac{1}{2\pi\beta\lambda} \exp\left(-\frac{|y_i|}{\lambda} - \frac{|\mathbf{b}_i|^2}{\beta}\right),
\end{equation}

\noindent with the widths in rapidity and impact parameter, 
$\lambda$ and $\beta$ respectively, being parameters to be determined.
Assuming that the partons interact independently with the target, the total interaction    can be written as
\begin{equation}
 t(\mathbf{b}_1,\dots,\mathbf{b}_N;y_1,\dots,y_N;\mathbf{b}) =  1-\prod_{i=1}^N(1-\tau(\mathbf{b}_i-\mathbf{b},y_i)),
\end{equation}

\noindent where $\mathbf{b}$ is the impact parameter of the
hadron-hadron collision and $\tau$ is the interaction probability 
for a single parton, considered to be given by
%
\begin{equation}
 \tau(\mathbf{b},y) = A \exp\left(-\frac{|y|}{\alpha} - \frac{|\mathbf{b}|^2}{\gamma}\right),
 \label{eq:MP_prob_interaction}
\end{equation}

\noindent with, again, $\alpha$, $\lambda$ and $A$ parameters to be determined.
Such assumptions imply that  the cross sections in the $\mathbf{b}$ 
space can be expressed in terms of the mean value and dispersion 
of the eigenvalues $t$, namely,
\begin{align}
& \frac{d\sigmatot}{d^2\mathbf{b}} = 2\langle t \rangle = \left[1-\exp\left(-\dfrac{G^2 A}{\beta\xi}\left(\dfrac{\alpha/\lambda}{1+\alpha/\lambda}\right)e^{-b^2/(\gamma+\beta)}\right)\right],\label{eq:dsigmatotd2b_MP}\\[10pt]
& \frac{d\sigmael}{d^2\mathbf{b}} = \langle t \rangle^2 = \left[1-\exp\left(-\dfrac{G^2 A}{\beta\xi}\left(\dfrac{\alpha/\lambda}{1+\alpha/\lambda}\right)e^{-b^2/(\gamma+\beta)}\right)\right]^2,\label{eq:dsigmaeld2b_MP}\\[10pt]
& \frac{d\sigmadiff}{d^2\mathbf{b}} = \langle t^2 \rangle-\langle t\rangle^2 = \exp\left(-2 \dfrac{G^2A}{\beta\xi}\left(\dfrac{\alpha/\lambda}{1+\alpha/\lambda}\right)e^{-b^2/(\gamma+\beta)}\right)\left[\exp\left(\dfrac{G^2A^2}{\beta\zeta}\left(\dfrac{\alpha/\lambda}{2+\alpha/\lambda}\right)e^{-2b^2/(\gamma+2\beta)}\right)-1\right], \label{eq:dsigmadiffd2b_MP}
\end{align}

\noindent where the averages are taken over the probability distribution of the beam partons,  $\xi = \beta^{-1} + \gamma^{-1}$ and $\zeta = \beta^{-1} + (\gamma/2)^{-1}$.
%
%
%
By integrating these equations over 
$\mathbf{b}$, we are able to calculate the energy dependence of $\sigmatot(s)$, 
$\sigmael(s)$ and $\sigmadiff(s)$.

The MP model implies that the cross sections are completely specified once we know the parameters $\lambda$, $\alpha$, $A$, $G^2$ and $\beta$. As in the original paper ~\cite{Miettinen:1978}, we will assume that  $\alpha/\lambda = 2$ and 
$\gamma/\beta = 2$. Moreover, we will consider $A=1$, 
in order to Eq.~\eqref{eq:MP_prob_interaction} account for the maximal 
probability allowed. Therefore, we are left with only two free 
parameters to be determined: $G^2$ and $\beta$.
Our strategy, similar to that used in Refs.~\cite{Miettinen:1978,Sapeta:2005},
is to determine these parameters from the experimental 
values of $\sigmatot$ and $\sigmael$ by means of the 
integrals of Eqs.~\eqref{eq:dsigmatotd2b_MP} and \eqref{eq:dsigmaeld2b_MP}.
Specifically, we considerably expand the analysis presented 
in~\cite{Sapeta:2005} by including the data from both $pp$ and $\bar{p}p$
scattering in the largest energy range considered so far. 
In particular, we will include the most recent experimental 
information obtained in the LHC (up to 13 TeV), 
and consider data at low energies ($\gtrsim$ 10 GeV).
Since $G^2$ and $\beta$ are determined simultaneously by
$\sigmatot$ and $\sigmael$, we selected for  each energy those 
data that have been obtained by the same 
experimental group, in order to minimize the effects of systematics
of different experiments. 
For this, we used the dataset provided by the 
Particle Data Group (PDG)\cite{pdg:2018}
and Refs.\cite{Antchev:2017b,Nemes_talk:2018,Antchev:2013a,Antchev:2013b}.
All selected data for $pp$ 
\cite{Czapek:1962,Blobel:1973,Breitenlohner:1963,Ayres:1976,Brick:1982,Bartenev:1974,Barish:1974,Amos:1985,Baksay:1978,Amaldi:1973,Firestone:1974,Dao:1972,Antchev:2017b,Nemes_talk:2018,Antchev:2013a,Antchev:2013b} 
and $\bar{p}p$ \cite{Ayres:1976,Amos:1985,Abe:1993s,Abe:1993u,Bozzo:1984,Avila:1999,Amos:1990}
scattering are displayed in Tables~\ref{tab:data_results_MP_pp} 
and~\ref{tab:data_results_MP_ppbar}, respectively, together with 
our results for $G^2$ and $\beta$. The errors in the experimental
data corresponds to statistical and systematic uncertainties 
added in quadrature and the errors in the parameters of the
model were obtained from error propagation from the data uncertainties.
Once we have determined $G^2$ and $\beta$ from the experimental information,
we can find analytical expressions that describe their energy dependence by fitting  the 
 $G^2$ and $\beta$ values. Specifically, we note that $G^2$ presents a energy
dependence very close to that showed by $\sigmatot$ data for $pp$ and $\bar{p}p$ scattering. 
For this reason, we consider an analytical expression inspired by Regge phenomenology:
\begin{equation}
 G^2(s) = a\left(\frac{s}{s_0}\right)^{-b} + c\left(\frac{s}{s_0}\right)^{d},
 \label{eq:par_G2}
\end{equation}

\noindent where $a$, $b$, $c$ and $d$ are free parameters and $s_0 = 1$~GeV$^2$ is fixed.
On the other hand, $\beta$ has a linear dependence on the $\ln s$ variable. Therefore,
we consider
\begin{equation}
 \beta(s) = \beta_0 + \beta_1 \ln \left(\frac{s}{s_0}\right),
 \label{eq:par_beta}
\end{equation}

\noindent with $\beta_0$ and $\beta_1$ to be determined from the fit.
For energies above 10 GeV, the values of $G^2$ and $\beta$ for $pp$ and $\bar{p}p$ 
are consistent within uncertainties, therefore we consider the data of both reactions
as being of the same. The results of the fits for $\sqrt{s_\text{min}}=10$~GeV are
\begin{equation}
 \begin{array}{l|l}
  a = 4.1  \pm 1.7   & \mbox{}\\
  b = 0.25 \pm 0.13  & \beta_0 = 0.1809 \pm 0.0078\text{ fm}^2\\
  c = 0.80 \pm 0.31  & \beta_1 = 0.0115 \pm 0.0011\text{ fm}^2\\
  d = 0.101\pm 0.023 & \mbox{}
 \end{array}
 \label{eq:res_G2_beta}
\end{equation}
The fit to $G^2$ points present a reduced $\chi^2$ 
of 0.836 for 31 degreed of freedom (d.o.f) 
while for the fit to $\beta$ we have 0.875 for 33 d.o.f.,
indicating that our choices for the parametrization 
describe successfully the energy dependence of the parameters of MP model.


\begin{table}[t]
 \centering
 \begin{tabular}{c|c|c|c|c|c}\hline\hline
  Energy (GeV) & Ref. & $\sigmatot$ (mb) & $\sigmael$ (mb) & $G^2$ & $\beta$ (fm$^2$)\\\hline
  6.77 & \cite{Czapek:1962} & 39.7  $\pm$ 1.5  & 8.3  $\pm$ 1.2   &  3.08 $\pm$ 0.80  &  0.208 $\pm$ 0.042  \\
  6.84 & \cite{Blobel:1973} & 38.9  $\pm$ 0.1  & 8.3  $\pm$ 0.2   &  3.19 $\pm$ 0.14  &  0.198 $\pm$ 0.0063 \\
  6.91 & \cite{Breitenlohner:1963} & 39.3  $\pm$ 0.8  & 8.8  $\pm$ 0.3   &  3.48 $\pm$ 0.26  &  0.188 $\pm$ 0.012  \\
  9.78 & \cite{Ayres:1976} & 38.14 $\pm$ 0.19 & 7.61 $\pm$ 0.29  &  2.85 $\pm$ 0.18  &  0.212 $\pm$ 0.010  \\
  11.54 & \cite{Ayres:1976} & 38.24 $\pm$ 0.19 & 7.41 $\pm$ 0.31 &  2.71 $\pm$ 0.19  &  0.220 $\pm$ 0.012  \\
  13.76 & \cite{Ayres:1976} & 38.39 $\pm$ 0.19 & 7.07 $\pm$ 0.35 &  2.50 $\pm$ 0.20  &  0.235 $\pm$ 0.014  \\
  16.66 & \cite{Brick:1982} & 38.47 $\pm$ 0.58 & 6.85 $\pm$ 0.24 &  2.37 $\pm$ 0.14  &  0.246 $\pm$ 0.013  \\
  16.83 & \cite{Bartenev:1974} & 38.62 $\pm$ 0.07 & 6.97 $\pm$ 0.11 &  2.423$\pm$ 0.060 &  0.2427$\pm$ 0.0047 \\
  18.17 & \cite{Ayres:1976} & 38.76 $\pm$ 0.19 & 7.06 $\pm$ 0.28 &  2.46 $\pm$ 0.15  &  0.241 $\pm$ 0.012  \\
  19.66 & \cite{Barish:1974} & 39.00 $\pm$ 1.00 & 6.92 $\pm$ 0.44 &  2.36 $\pm$ 0.25  &  0.250 $\pm$ 0.024  \\
  23.5  & \cite{Amos:1985} & 39.65 $\pm$ 0.23 & 6.81 $\pm$ 0.33 &  2.25 $\pm$ 0.17  &  0.264 $\pm$ 0.016  \\
  23.5  & \cite{Baksay:1978} & 39.13 $\pm$ 0.40 & 6.82 $\pm$ 0.08 &  2.297$\pm$ 0.054 &  0.2564$\pm$ 0.0066 \\
  23.5  & \cite{Amaldi:1973} & 38.90 $\pm$ 0.70 & 6.70 $\pm$ 0.30 &  2.26 $\pm$ 0.16  &  0.259 $\pm$ 0.017  \\
  23.76 & \cite{Firestone:1974} & 40.68 $\pm$ 0.55 & 7.89 $\pm$ 0.52 &  2.72 $\pm$ 0.30  &  0.234 $\pm$ 0.020  \\
  23.88 & \cite{Dao:1972} & 39.00 $\pm$ 1.00 & 7.20 $\pm$ 0.40 &  2.51 $\pm$ 0.24  &  0.238 $\pm$ 0.021  \\
  30.6  & \cite{Amos:1985} & 40.11 $\pm$ 0.19 & 6.75 $\pm$ 0.32 &  2.18 $\pm$ 0.15  &  0.274 $\pm$ 0.016  \\
  30.6  & \cite{Baksay:1978} & 39.91 $\pm$ 0.41 & 7.39 $\pm$ 0.08 &  2.522$\pm$ 0.059 &  0.2431$\pm$ 0.0064 \\
  30.6  & \cite{Amaldi:1973} & 40.20 $\pm$ 0.80 & 6.90 $\pm$ 0.40 &  2.24 $\pm$ 0.21  &  0.268 $\pm$ 0.022  \\
  44.9  & \cite{Baksay:1978} & 41.89 $\pm$ 0.41 & 7.45 $\pm$ 0.08 &  2.368$\pm$ 0.053 &  0.2679$\pm$ 0.0067 \\
  52.8  & \cite{Amos:1985} & 42.38 $\pm$ 0.17 & 7.17 $\pm$ 0.30 &  2.20 $\pm$ 0.14  &  0.288 $\pm$ 0.015  \\
  52.8  & \cite{Baksay:1978} & 42.85 $\pm$ 0.42 & 7.56 $\pm$ 0.08 &  2.340$\pm$ 0.051 &  0.2767$\pm$ 0.0069 \\
  62.3  & \cite{Amos:1985} & 43.55 $\pm$ 0.32 & 7.51 $\pm$ 0.36 &  2.26 $\pm$ 0.16  &  0.289 $\pm$ 0.017  \\
  62.5  & \cite{Baksay:1978} & 44.00 $\pm$ 0.45 & 7.77 $\pm$ 0.10 &  2.343$\pm$ 0.059 &  0.2838$\pm$ 0.0077 \\
  2760  & \cite{Antchev:2017b,Nemes_talk:2018} & 84.7  $\pm$  3.3 & 21.8 $\pm$ 1.4  &  4.62 $\pm$ 0.77  &  0.335 $\pm$ 0.044  \\
  7000  & \cite{Antchev:2013a} & 98.0  $\pm$  2.5 & 25.1 $\pm$ 1.1  &  4.57 $\pm$ 0.51  &  0.390 $\pm$ 0.035  \\
  8000  & \cite{Antchev:2013b} & 101.7 $\pm$  2.9 & 27.1 $\pm$ 1.4  &  4.99 $\pm$ 0.69  &  0.382 $\pm$ 0.040  \\
  13000 & \cite{Antchev:2017b} & 110.6 $\pm$  3.4 & 31.0 $\pm$ 1.7  &  5.66 $\pm$ 0.92  &  0.385 $\pm$ 0.045  \\
\hline\hline
 \end{tabular}
  \caption{\label{tab:data_results_MP_pp}Experimental data on $\sigmatot$ and $\sigmael$
 from $pp$ scattering considered in this analysis and the values of the parameters 
 $G^2$ and $\beta$ of Miettinen-Pumplin model determined with Eqs.~\eqref{eq:dsigmatotd2b_MP} and 
 \eqref{eq:dsigmaeld2b_MP}. References of experimental data are also shown.}
\end{table}

\begin{table}[ht!]
 \centering
  \begin{tabular}{c|c|c|c|c|c}\hline\hline
  Energy (GeV) & Ref. & $\sigmatot$ (mb) & $\sigmael$ (mb) & $G^2$ & $\beta$ (fm$^2$)\\\hline
  9.78  & \cite{Ayres:1976} & 43.86 $\pm$ 0.22 & 8.2   $\pm$ 0.4  & 2.56 $\pm$ 0.20  & 0.264 $\pm$ 0.016 \\
  11.54 & \cite{Ayres:1976} & 43.00 $\pm$ 0.22 & 7.30  $\pm$ 0.47 & 2.21 $\pm$ 0.21  & 0.291 $\pm$ 0.022 \\
  13.76 & \cite{Ayres:1976} & 42.04 $\pm$ 0.21 & 7.8   $\pm$ 0.6  & 2.53 $\pm$ 0.31  & 0.255 $\pm$ 0.024 \\
  16.26 & \cite{Ayres:1976} & 41.80 $\pm$ 0.21 & 7.52  $\pm$ 0.6  & 2.41 $\pm$ 0.30  & 0.264 $\pm$ 0.026 \\
  18.17 & \cite{Ayres:1976} & 41.60 $\pm$ 0.21 & 7.12  $\pm$ 0.52 & 2.23 $\pm$ 0.25  & 0.279 $\pm$ 0.024 \\
  30.4  & \cite{Amos:1985} & 42.13 $\pm$ 0.58 & 7.16  $\pm$ 0.44 & 2.21 $\pm$ 0.21  & 0.284 $\pm$ 0.023 \\
  52.6  & \cite{Amos:1985} & 43.32 $\pm$ 0.34 & 7.44  $\pm$ 0.44 & 2.25 $\pm$ 0.20  & 0.289 $\pm$ 0.021 \\
  62.3  & \cite{Amos:1985} & 44.12 $\pm$ 0.40 & 7.46  $\pm$ 0.44 & 2.19 $\pm$ 0.19  & 0.300 $\pm$ 0.022 \\
  546   & \cite{Abe:1993s,Abe:1993u} & 61.26 $\pm$ 0.93 & 12.87 $\pm$ 0.30 & 3.11 $\pm$ 0.15  & 0.319 $\pm$ 0.015 \\
  547   & \cite{Bozzo:1984} & 61.90 $\pm$ 1.6  & 13.30 $\pm$ 0.61 & 3.23 $\pm$ 0.31  & 0.313 $\pm$ 0.027 \\
  1800  & \cite{Avila:1999} & 71.71 $\pm$ 2.02 & 15.79 $\pm$ 0.87 & 3.38 $\pm$ 0.38  & 0.351 $\pm$ 0.034 \\
  1800  & \cite{Abe:1993s,Abe:1993u} & 80.03 $\pm$ 2.24 & 19.70 $\pm$ 0.85 & 4.20 $\pm$ 0.45  & 0.337 $\pm$ 0.030 \\
  1800  & \cite{Amos:1990} & 72.10 $\pm$ 3.3  & 16.6  $\pm$ 1.6  & 3.67 $\pm$ 0.75  & 0.333 $\pm$ 0.056 \\
\hline\hline
 \end{tabular}
 \caption{\label{tab:data_results_MP_ppbar}Experimental data on $\sigmatot$ and $\sigmael$
 from $\bar{p}p$ scattering considered in this analysis and the values of the parameters 
 $G^2$ and $\beta$ of Miettinen-Pumplin model determined with Eqs.~\eqref{eq:dsigmatotd2b_MP} and 
 \eqref{eq:dsigmaeld2b_MP}. References of experimental data are also shown.}
\end{table}

\begin{figure}[t]
 \centering
 \includegraphics[width=0.31\textwidth]{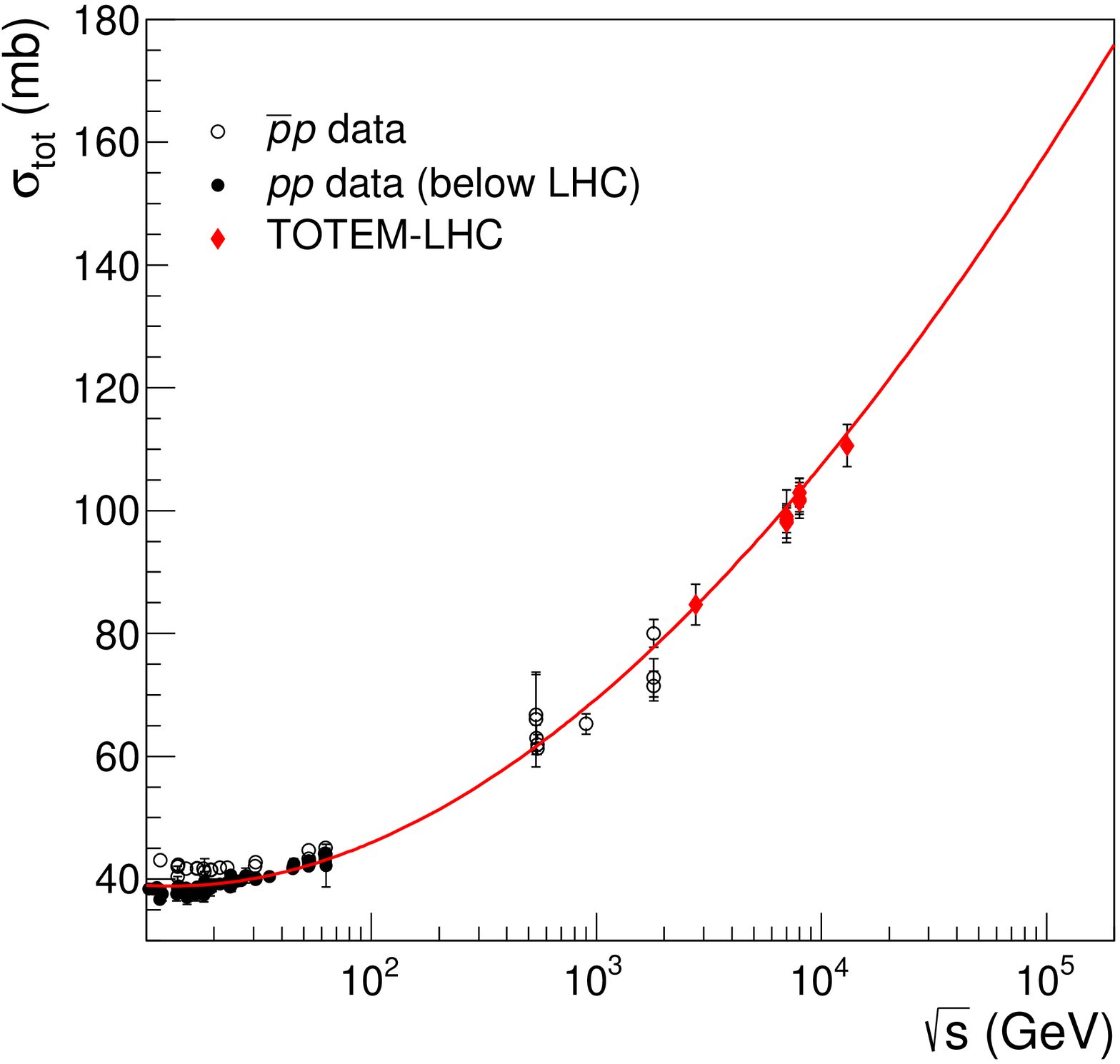}
 \includegraphics[width=0.31\textwidth]{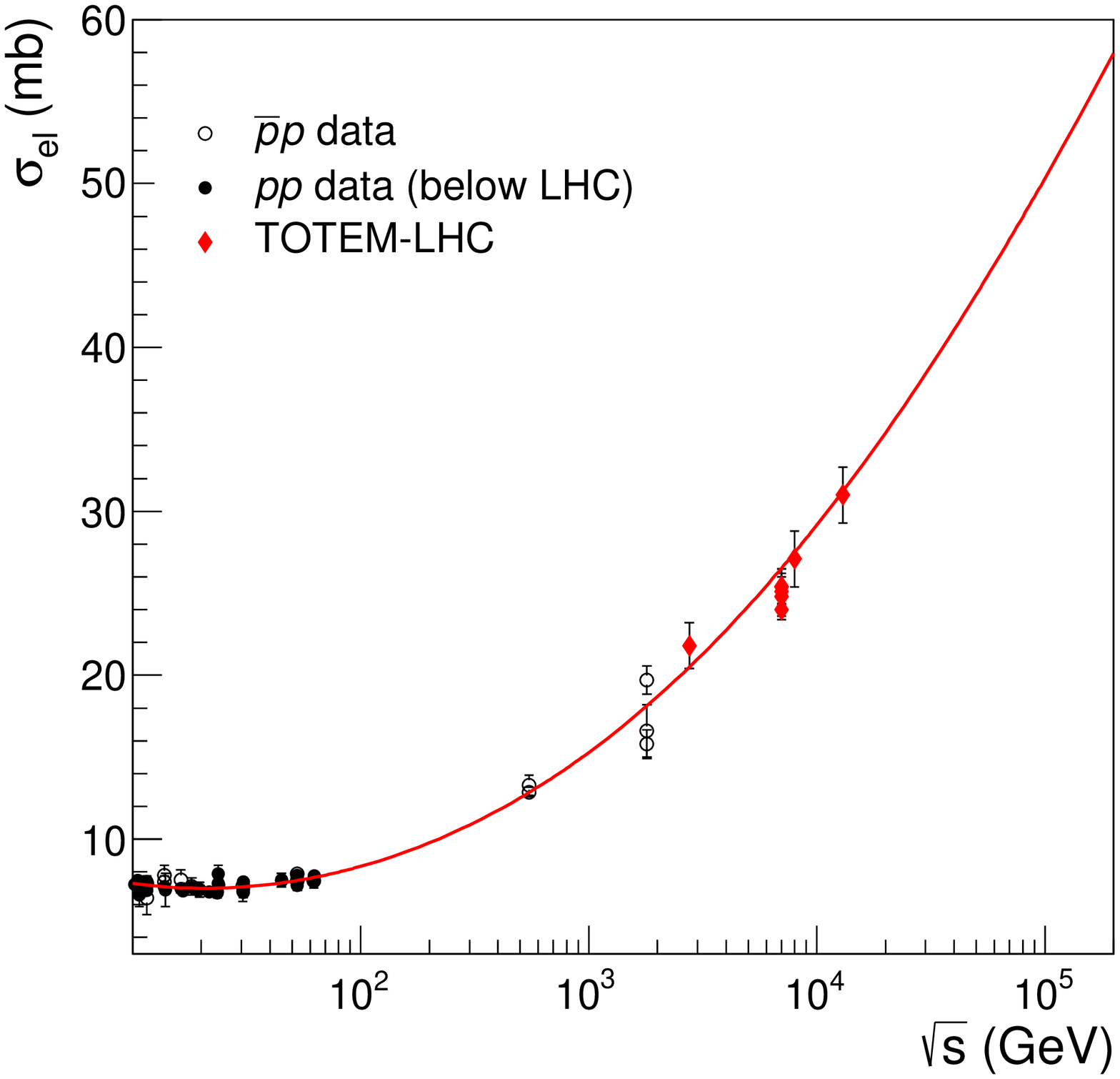}
 \includegraphics[width=0.31\textwidth]{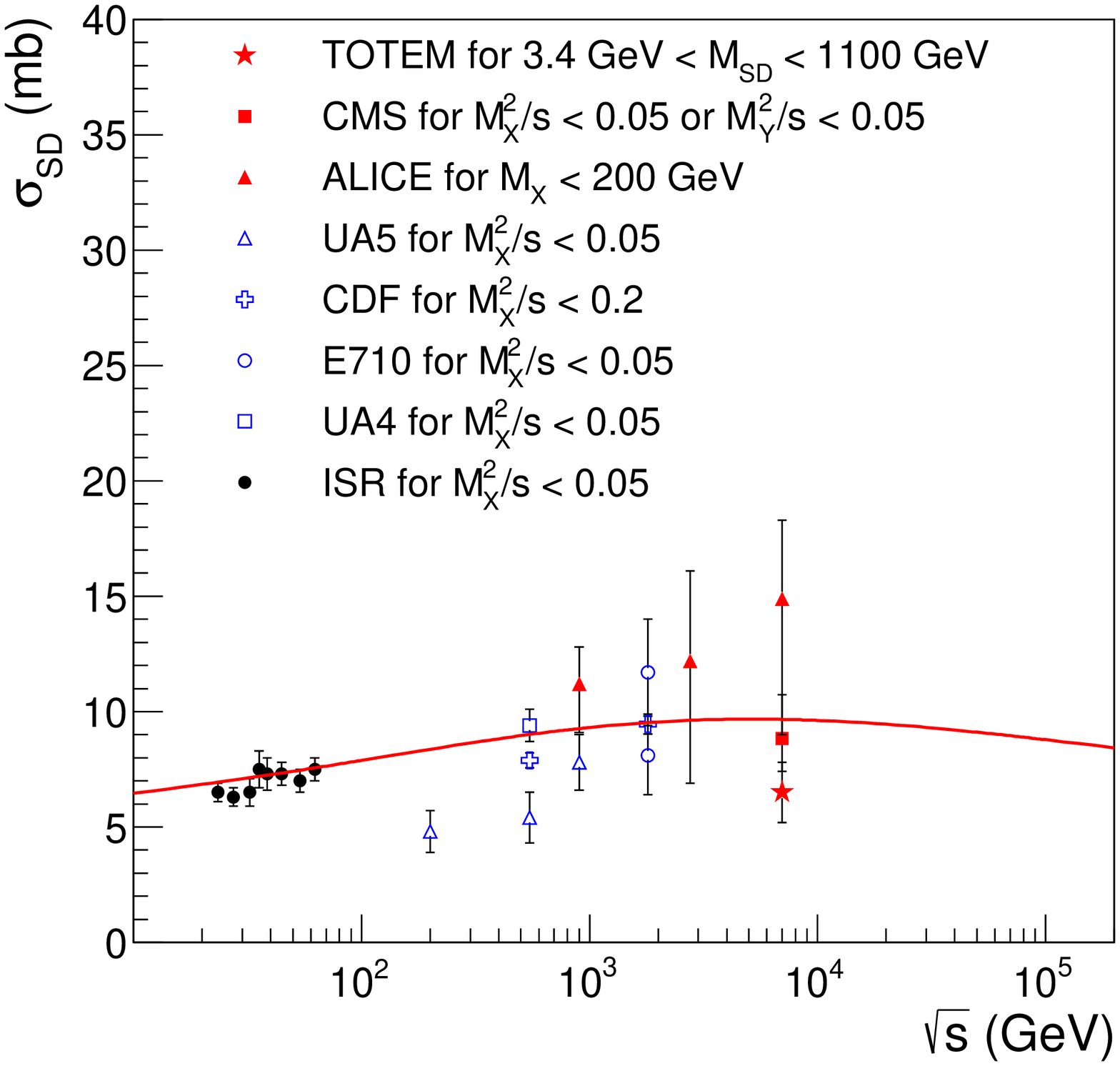}
 \caption{\label{fig:cross_sections_pp_MP} Total (left), elastic (central) and single diffractive dissociation
 (right) cross sections as a function of energy, calculated from Eqs.~\eqref{eq:dsigmatotd2b_MP}-\eqref{eq:res_G2_beta}, and comparison with the available experimental data.}
\end{figure}

Using our parametrizations for $G^2$ and $\beta$  we are able to calculate the energy dependence of $pp$ and $\bar{p}p$ cross sections. The results are presented in Fig.~\ref{fig:cross_sections_pp_MP}, where we compare our predictions with the available data for $\sigmatot$, $\sigmael$ \cite{pdg:2018,Antchev:2013c,Antchev:2013b,Antchev:2015,Antchev:2016,Antchev:2017b,Nemes_talk:2018} and
for single dissociation (SD) cross section \cite{Armitage:1982,Cartiglia:2013,alice:2013,cms:2015,ua5:1986a,Alner:1987,Bernard:1987,Amos:1990,Amos:1993,cdf:1994a}. 
We have that the MP model, is able to describe the current experimental data. In particular, the experimental data for the diffractive cross section, which was not considered in the fitting of the parameters, is quite well described. In Table~\ref{tab:MP_pred}, 
we summarize the predictions for these cross sections at energies of 
the LHC and cosmic rays experiments. The values of $\sigmatot$ and $\sigmael$ derived using the MP model are in agreement with predictions of other approaches. 
For instance, the $\sigmatot$ from the analysis performed in Ref.~\cite{Fagundes:2017}
are 112.1 $\pm$ 1.3~mb at 14 TeV, 143.4 $\pm$ 2.8~mb at 57 TeV 
and 156.2 $\pm$ 3.6~mb at 95 TeV, while for $\sigmael$ (considering a combination 
of these predictions with those obtained in \cite{Fagundes:2016}) 
we have 31.05 $\pm$ 0.67~mb, 45.0 $\pm$ 1.3~mb and 
51.2 $\pm$ 1.6~mb at 14, 57 and 95 TeV, respectively. For the diffractive cross section, we predict that the cross section for $\sqrt{s} = 14$ TeV will be almost identical to that measured in the Run I of the LHC. Finally, 
it is also interesting to note that the total cross section for $pp$ scattering in the MP model behaves assimptotically as $\sigmatot \sim \beta \ln G^2$. Therefore, with our choices of $G^2$ and $\beta$, we have
$\sigmatot \sim \ln^2 s$, in accordance with the Froissart-Martin bound~\cite{Froissart:1961,Martin:1965}.

\begin{table}[ht!]
\centering
\begin{tabular}{c|c|c|c|c}\hline\hline
Energy            & 8 TeV & 14 TeV & 57 TeV & 95 TeV \\\hline
$\sigmatot$ (mb)  & 103.1 & 114.1  & 144.8  & 157.1  \\
$\sigmael$ (mb)   & 27.5  & 31.8   & 44.5   & 49.8   \\
$\sigmadiff$ (mb) & 9.7   & 9.6    & 9.0    & 8.8    \\\hline\hline
\end{tabular}
\caption{\label{tab:MP_pred}Predictions to $\sigmatot$, $\sigmael$ and $\sigmadiff$ at energies of the LHC and 
of cosmic rays experiments obtained from Miettinen-Pumplin model with the energy dependence
of parameters $G^2$ and $\beta$ given by Eqs.~\eqref{eq:par_G2}-\eqref{eq:res_G2_beta}.}
\end{table}

\begin{figure}[t]
 \centering
 \includegraphics[scale=0.29]{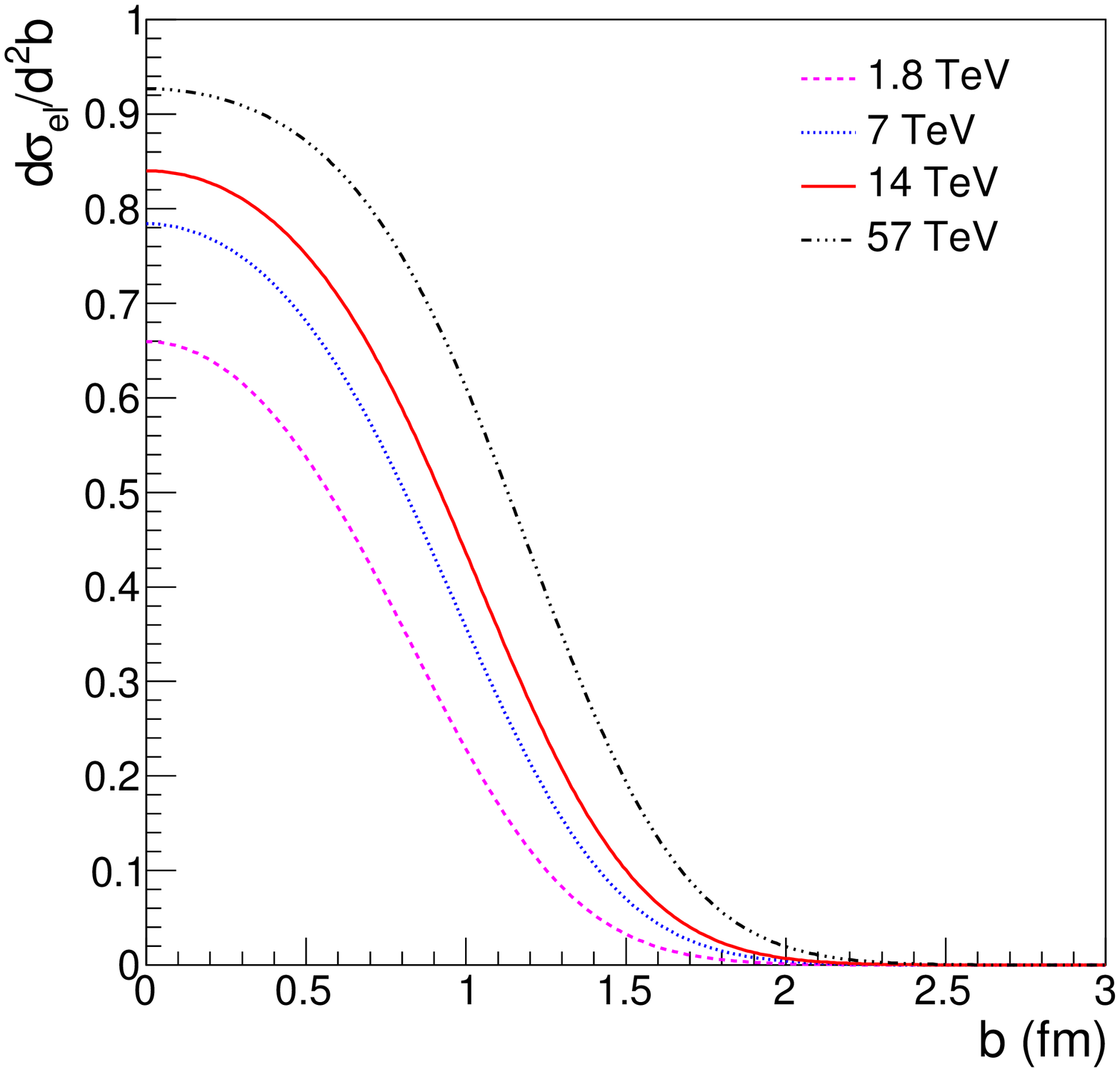}
 \includegraphics[scale=0.29]{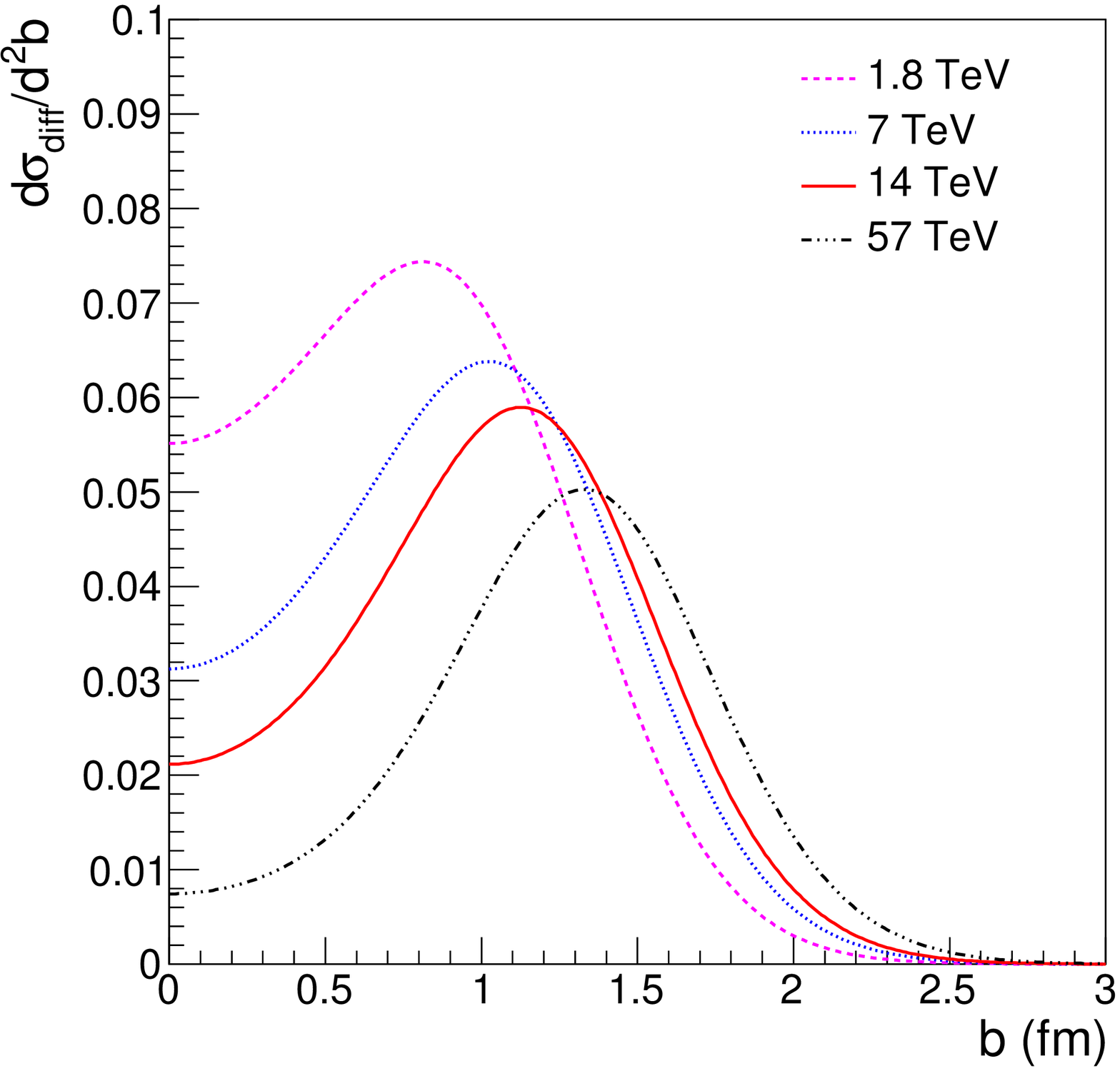}
 \includegraphics[scale=0.29]{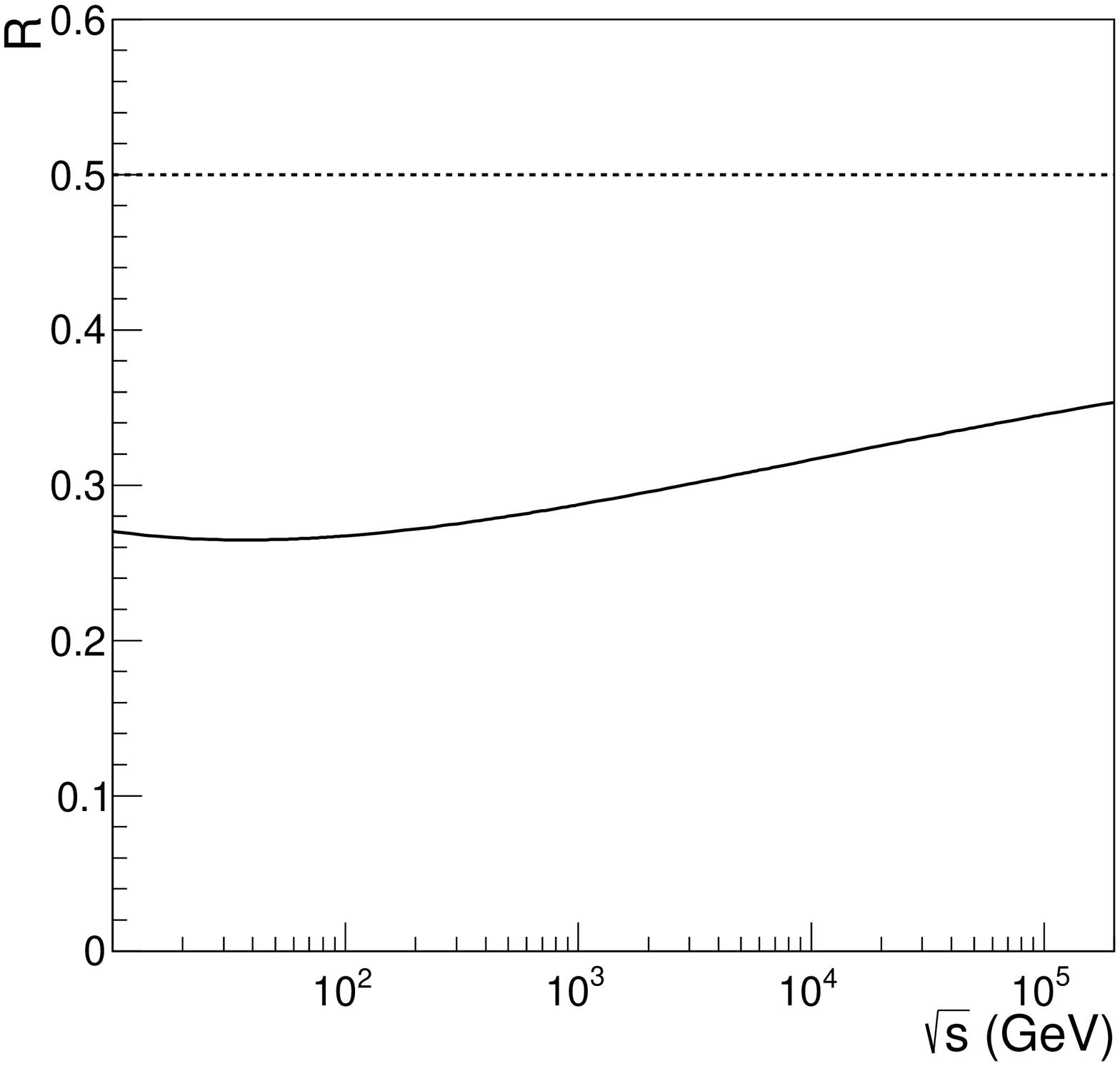}
 \caption{\label{fig:cross_sections_bspace} Energy evolution of the elastic and diffractive dissociation cross sections in the impact parameter space, left and central panels respectively, and of the ratio $R = (\sigmael + \sigmadiff)/\sigmatot$ (right panel).}
\end{figure}

One of the important characteristics of the MP model is that it  also allow us to study the energy evolution of the cross sections in the impact parameter space. 
The elastic and diffractive differential cross section
as a function of $b$ are presented  in the left and central panels of 
Fig.~\ref{fig:cross_sections_bspace}, respectively, 
from Tevatron to  cosmic rays energies.
We observe that the elastic scattering is mainly central and 
its magnitude increases with the energy, approaching the black disc limit,
although it is not yet saturated at 57 TeV. In contrast, the diffractive dissociation becomes more peripheral 
with its maximum moving to larger impact parameter as the energy increases. We also note the decrease of the magnitude of this cross section from lower to higher energies. Finally, in the right panel of Fig.~\ref{fig:cross_sections_bspace}, we present the evolution with energy
of the ratio $R(s) = (\sigmael+\sigmadiff)/\sigmatot$, where the cross sections are those of 
Eqs.~\eqref{eq:dsigmatotd2b_MP}-\eqref{eq:dsigmadiffd2b_MP} integrated in $\mathbf{b}$-space. 
According to the Pumplin bound~\cite{Pumplin:1973}, we should have $R\leq 1/2$. Our results indicate that
this bound is not saturated and that $R$ approaches the upper limit in a slow rate.

\section{Diffractive excitation in $pA$ collisions}
\label{sec:difpA}

Our goal in this Section is to extend our previous analysis for proton -- nucleus collisions. The extrapolation of results from $pp$ to $pA$ collisions is generally performed using the Glauber formalim \cite{Glauber:1970}, which assumes that the projectile nucleon travel along straight lines undergoing multiple elastic collisions with the nucleons in the target. As pointed by Gribov \cite{gribov}, such formalism disregard the diffractive excitation of the intermediate nucleons, which gives a significant contribution for the total and elastic $pA$ cross sections. We will include the contribution of the diffractive excitation using the formalism formulated in Refs. \cite{Heiselberg:1991is,Blaettel:1993}, which is based on the Good - Walker approach (For a review see Ref. \cite{Frankfurt:2011cs}). As in the GW approach,  the incoming hadron state is expressed as a superposition of the eigenstates of the scattering operator. The basic assumption in Refs. \cite{Heiselberg:1991is,Blaettel:1993} is that each of these eigenstates interacts with the target with its own cross section $\sigma$, with the probability of interaction being given by $P(\sigma,s)$.  As a consequence, 
the total, elastic and diffractive $pA$ cross sections can be expressed as follows
\begin{align}
 \sigmatot^{pA}(s) & = 2\int d\sigma P(\sigma,s)\int d^2\mathbf{b} \Real \Gamma_A(\mathbf{b},\sigma), \label{eq:sigtot_pA_GS}\\
 \sigmael^{pA}(s) & = \int d\sigma \left|\int d^2\mathbf{b} P(\sigma,s) \Gamma_A(\mathbf{b},\sigma)\right|^2, \label{eq:sigel_pA_GS}\\
 \sigmadiff^{pA}(s) & = \int d^2\mathbf{b} \left[ \int d\sigma P(\sigma, s)|\Gamma_A(\mathbf{b},\sigma)|^2 - \left|\int d\sigma  P(\sigma,s) \Gamma_A(\mathbf{b},\sigma)\right|^2\right] \label{eq:sigdiff_pA_GS}.
\end{align}

\noindent where $\Gamma_A$ is the profile function for the $pA$ scattering, given by
\begin{equation}
 \Gamma_A(\mathbf{b},\sigma) = 1 - \exp\left(-\frac{A}{2}\sigma(1-i\eta) T(\mathbf{b})\right).
\end{equation}

\noindent Here $\eta$ is the ratio of the real to the 
imaginary part of the forward $pp$ elastic scattering amplitude
(which, given that it is small at high-energies, 
we shall consider $\eta\approx 0$), and $T(b)$ is 
the thickness function of the nucleus
\begin{equation}
 T(\mathbf{b}) = \frac{1}{2\pi B(s)}\int dz d^2\mathbf{s}\, e^{-(\mathbf{b}-\mathbf{s})^2/(2B(s))}\rho_A(\sqrt{|\mathbf{s}|^2 + z^2})
\end{equation}

\noindent with $\mathbf{b}$ being the impact parameter 
of the $pA$ scattering, $\mathbf{s}$ a vector in the 
$\mathbf{b}$-plane, $B(s)$ the forward slope of the 
$pp$ differential elastic cross section and 
$\rho_A$ the single nucleon distribution.
For heavy nucleus, we will consider the Wood-Saxon  distribution
\begin{equation}
 \rho_A (r) = \frac{\rho_0}{1+\exp((r-R_0)/a)},
 \label{eq:dist_nucleon}
\end{equation}

\noindent where $\rho_0$ is a normalization constant 
(calculated to have $\rho$ normalized to 1)
and the parameters $R_0$ and $a$ are available in Ref.~\cite{Loizides:2014}.

The main input in the Eqs. (\ref{eq:sigtot_pA_GS}), (\ref{eq:sigel_pA_GS}) and (\ref{eq:sigdiff_pA_GS}) is   the probability distribution $P(\sigma,s)$, which  is parametrized by  \cite{Heiselberg:1991is,Blaettel:1993}
\begin{equation}
 P(\sigma,s) = N(s)\frac{\sigma}{\sigma + \sigma_0(s)}\exp\left(-\frac{(\sigma/\sigma_0(s)-1)^2}{\Omega^2(s)}\right)\,\,,
\end{equation}

\noindent  
where the parameters $N$, $\sigma_0$ and $\Omega$ are energy dependent and 
are constrained by the moments of the distribution, which are assumed to satisfy the following properties
\begin{align}
 & \int P(\sigma,s)d\sigma  = 1,\\
 & \int \sigma P(\sigma,s)d\sigma =  \sigmatot(s),\\
 & \int \sigma^2 P(\sigma,s)d\sigma = \sigmatot^2(1+\omega_\sigma(s)),
\end{align}


\noindent where $\sigmatot$ is the total $pp/\bar{p}p$ cross section and $\omega_\sigma$, 
which is related to the broadness of the cross section fluctuations, is expressed in terms of the diffractive and elastic $pp/\bar{p}p$ cross sections as follows
 \cite{Blaettel:1993}
\begin{equation}
 \omega_\sigma = \frac{d\sigmadiff/dq^2|_{q^2=0}}{d\sigmael/dq^2|_{q^2=0}}.
\end{equation}

In what follows, we will use the results derived in the previous Section, obtained with the Miettinen-Pumplin model,  to constraint the parameters present in $P(\sigma,s)$ and to  calculate $\omega_\sigma$.
Moreover, the intrinsic preditive power of MP model that follows from our
parametrizations for $G^2(s)$ and $\beta(s)$, Eqs.~\eqref{eq:par_G2}-\eqref{eq:res_G2_beta},
allows to calculate of $\omega_\sigma$ for any desirable energy, and consequently, to estimate the energy dependence of the total, elastic and diffractive $pA$ cross sections. In fact, from the MP model, we have
\begin{align}
 \left.\frac{d\sigmael}{dq^2}\right\rvert_{q^2=0} & = \frac{1}{4\pi} \left[\int d^2\mathbf{b} \left(1 - e^{-G^2\langle\tau(\mathbf{b})\rangle}\right)\right]^2, \label{eq:dsdt_el_forward}\\[10pt]
 \left.\frac{d\sigmadiff}{dq^2}\right\rvert_{q^2=0} & = \frac{1}{4\pi}\int d^2\mathbf{b}\,d^2\mathbf{b}'\exp\left\{-G^2\left[\langle\tau(\mathbf{b})\rangle+\langle\tau(\mathbf{b}')\rangle\right]\right\}\left\{\exp\left[G^2\langle\tau(\mathbf{b})\tau(\mathbf{b}')\rangle\right]-1\right\}, \label{eq:dsdt_diff_forward}
\end{align}

\noindent where
\begin{align}
 & \langle\tau(\mathbf{b})\rangle  = \frac{A}{\beta\xi}\left(\frac{\alpha/\lambda}{1+\alpha/\lambda}\right) e^{-b^2/(\gamma+\beta)} \label{eq:tau_avarage}\\ 
 & \langle\tau(\mathbf{b})\tau(\mathbf{b}')\rangle  = \frac{A\gamma}{\gamma+2\beta}\left(\frac{\alpha/\lambda}{2+\alpha/\lambda}\right)\exp\left\{-\eta(b^2 + b'^2) + \mu b b'\cos(\phi-\phi') \right\}, \label{eq:tautau_avarage}
\end{align}

\begin{figure}[t]
 \centering
 \includegraphics[scale=0.4]{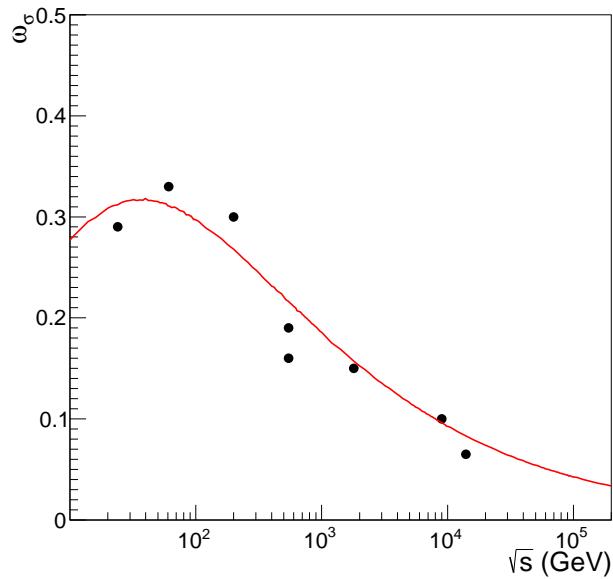}
 \caption{\label{fig:omega_sigma} Energy evolution of $\omega_\sigma$ 
 from the Miettinen-Pumplin model (red solid curve) compared with the 
 values presented in Ref.~\cite{Guzey:2006} (black dots).}
\end{figure}

\noindent $\eta = (\gamma+\beta)/(\gamma^2+2\gamma\beta)$, $\mu = 2\beta/(\gamma^2+2\gamma\beta)$, $b = |\mathbf{b}|$ and $\phi$ is the polar angle of $\mathbf{b}$ (analogous for $b'$ and $\phi'$). Using the parameters constrained in the previous Section, the resulting energy dependence of  $\omega_\sigma$  is presented in Fig.~\ref{fig:omega_sigma}
together with some values for this quantity estimated  
by Guzey and Strikman (GS) in Ref. ~\cite{Guzey:2006}. 
We have that our predictions are, in the average, similar to those presented in 
Ref. ~\cite{Guzey:2006}, as well the behavior predicted for the energy dependence. At low-energies, $\omega_\sigma$ rises with energy,
as expected from Regge phenomelogy \cite{Blaettel:1993} and  at high-energies  it decreases 
and tends to zero. Therefore, asymptotically the cross section fluctuations ceases, which is expected to occur in the black disc limit.
 
\begin{figure}[t]
 \centering
 \includegraphics[width=0.4\textwidth]{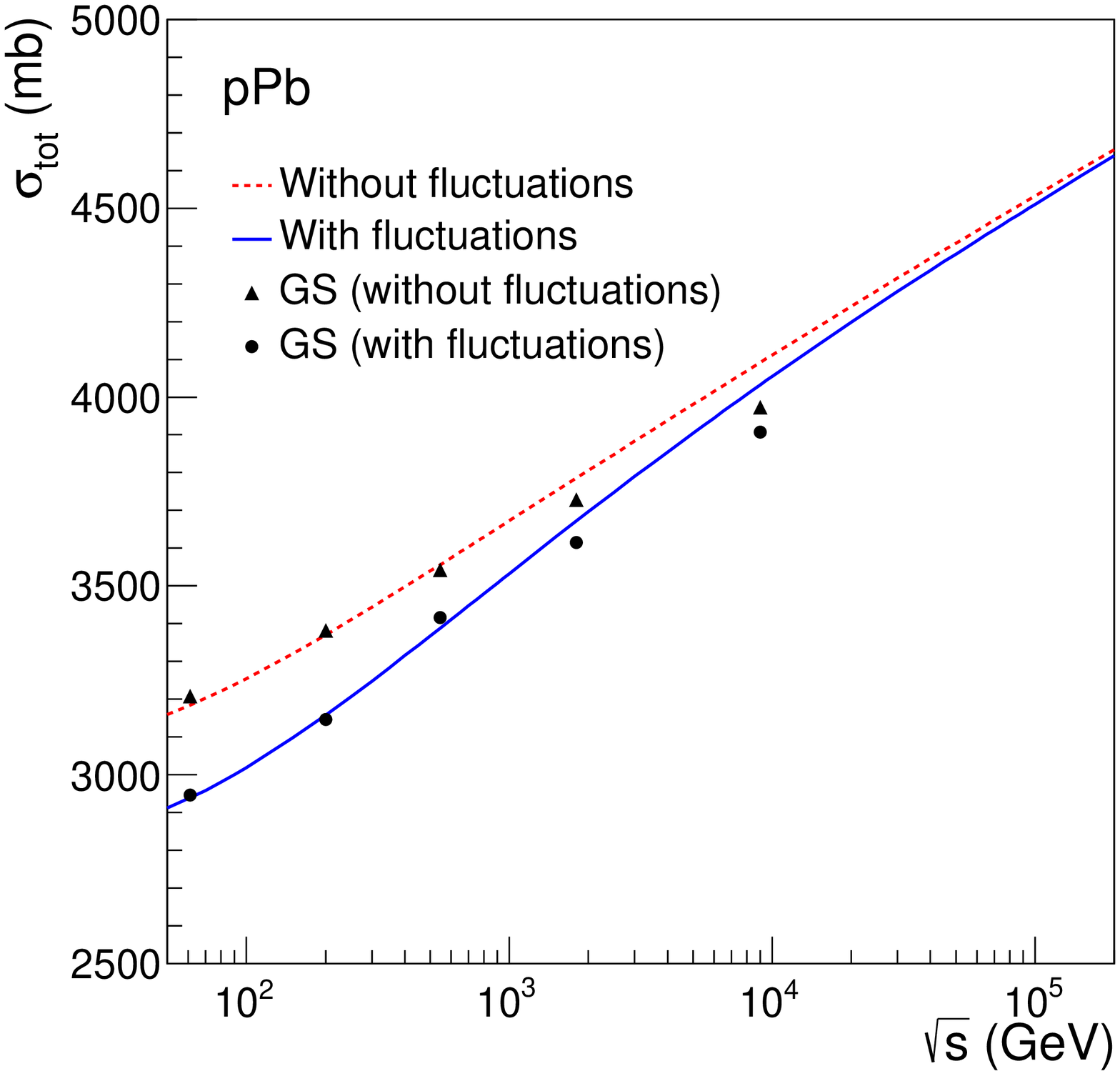}
 \includegraphics[width=0.4\textwidth]{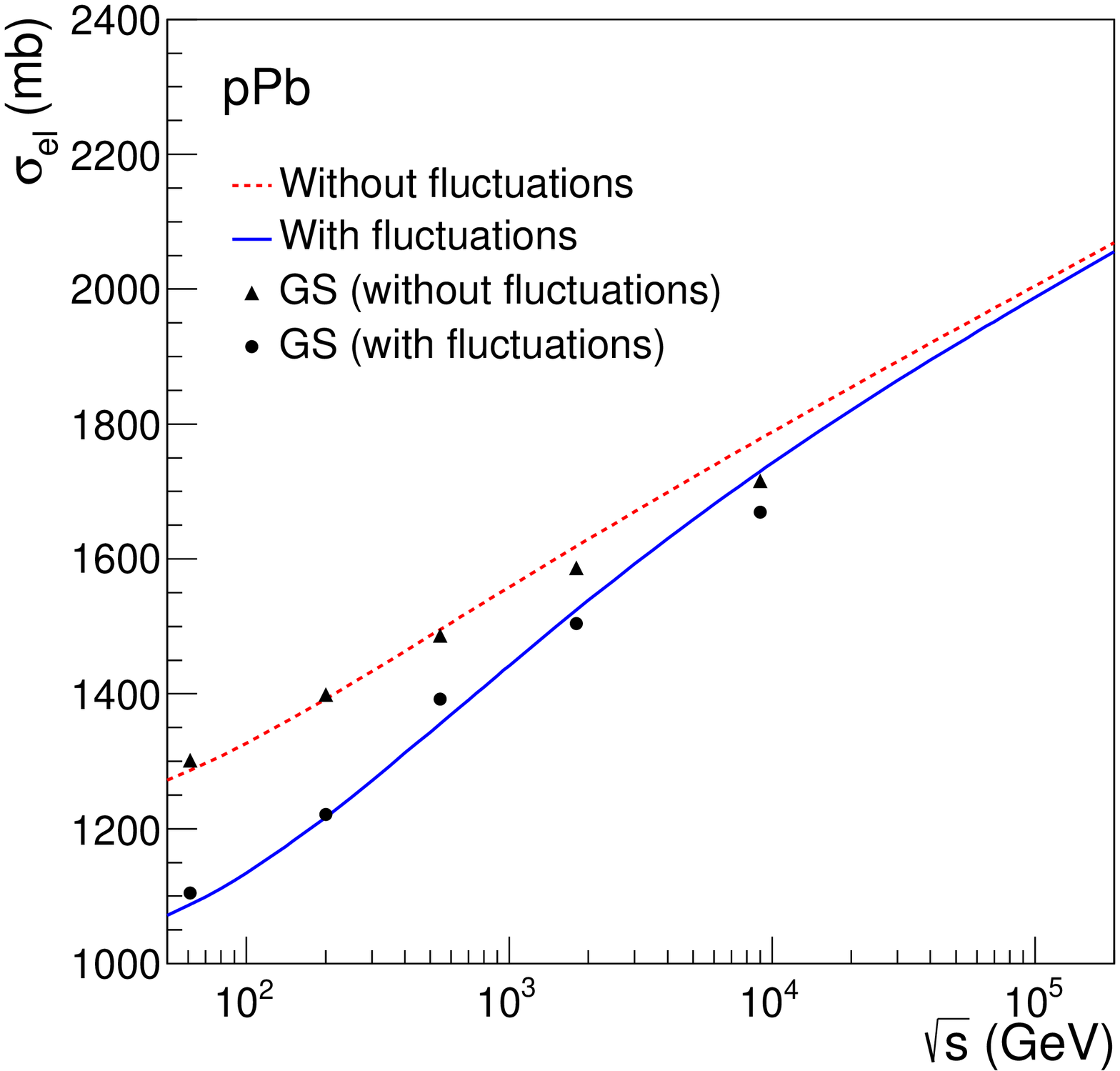}\\
 \includegraphics[width=0.4\textwidth]{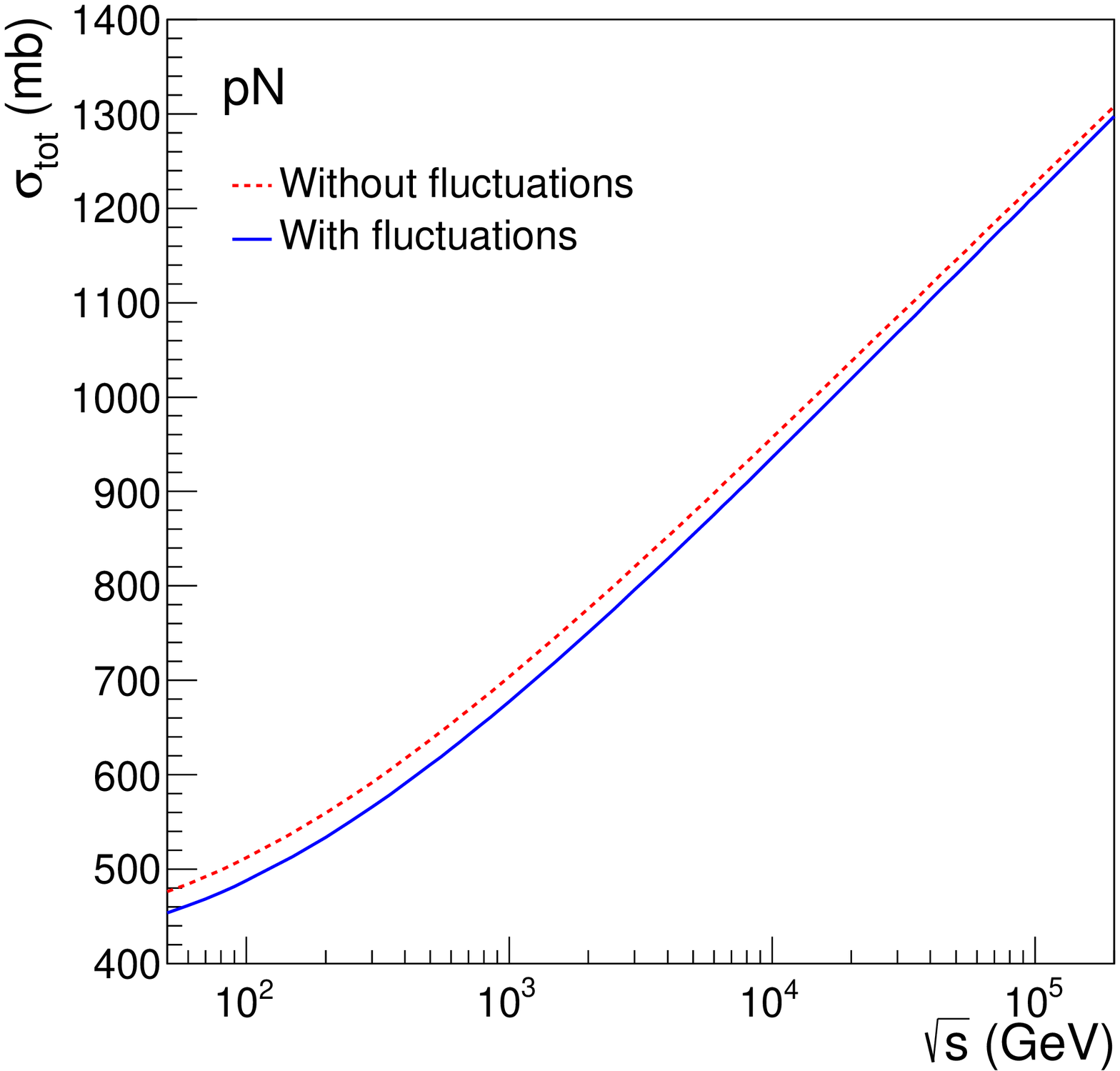}
 \includegraphics[width=0.4\textwidth]{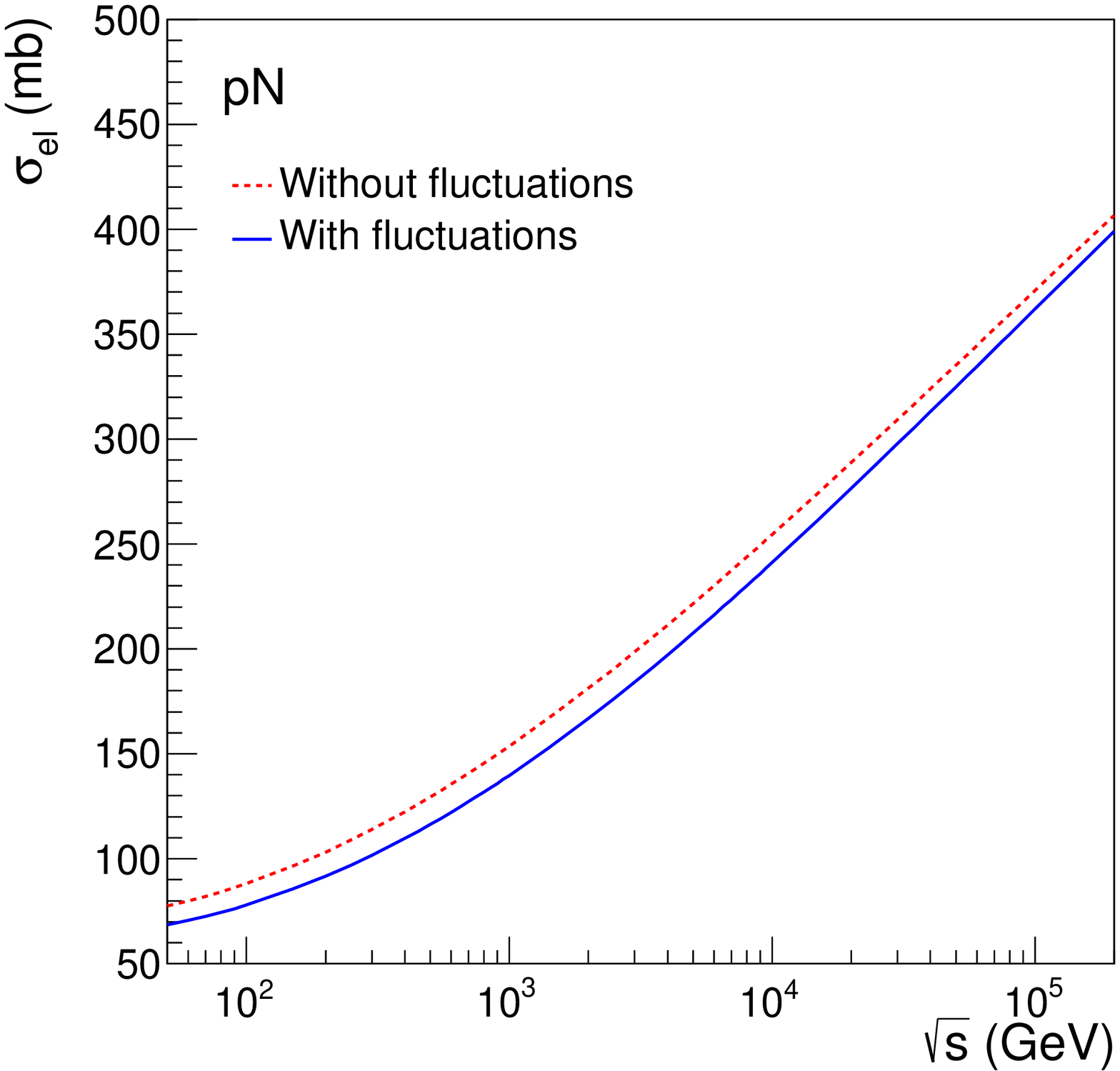}
 \caption{\label{fig:sigma_tot_el_pA} Total (left) and elastic (right)
 cross section for proton -- lead (top) and proton -- nitrogen (bottom) scattering with and 
 without cross section fluctuations as a function of the energy 
 in the center of mass system. In the $pPb$ case, 
 we also present the results by Guzey and Strikman (GS)~\cite{Guzey:2006}.}
\end{figure}

In order to estimate the $pA$ cross sections we will consider the predictions for the total $pp/\bar{p}p$ cross sections derived using the MP model. Moreover, we will assume that the slope of the amplitude $B(s)$ is given by
\begin{equation}
 B(s) = 11.21 - 0.176 \ln(s/s_0) + 0.0372 \ln^2 (s/s_0),
\end{equation}

\noindent where the parameters are given in GeV$^{-2}$ and have been constrained using the recent LHC data. Moreover, we assume  $s_0=1$~GeV$^{2}$. In Fig. ~\ref{fig:sigma_tot_el_pA} we present our predictions for the energy dependence of the  total and elastic cross sections  considering  proton -- lead ($pPb$) and proton -- nitrogen ($pN$) collisions\footnote{For the nitrogen nucleus, we considered a gaussian distribution 
for $\int \rho_A(\mathbf{s},z)dz$.}. 
For comparison, we also present the results obtained 
without cross section fluctuations, i.e. using the standard Glauber formalism. Moreover,  for $pPb$ collisions, we also show the results reported in Ref.~\cite{Guzey:2006}.
We see that the effects of the fluctuations is 
stronger at lower energies and heavier nuclei. 
As already expected from the  results for $\omega_\sigma$, the impact of the fluctuations become less important 
at high-energies. Our predictions for $pPb$ collisions are similar to those derived in Ref. ~\cite{Guzey:2006}, mainly at smaller energies. The difference between the predictions at higher energies  arises from the different inputs
used to constrain the probability parameters. 
Here we considered $\sigmatot$ from the MP model and 
$B(s)\sim \ln^2 s$, while in Ref. ~\cite{Guzey:2006} the authors have used $\sigmatot$ from a fit by Donnachie and Landshoff \cite{Donnachie:1992} and assumed $B(s)\sim \ln s$.

\begin{figure}[t]
 \centering
 \includegraphics[width=0.31\textwidth]{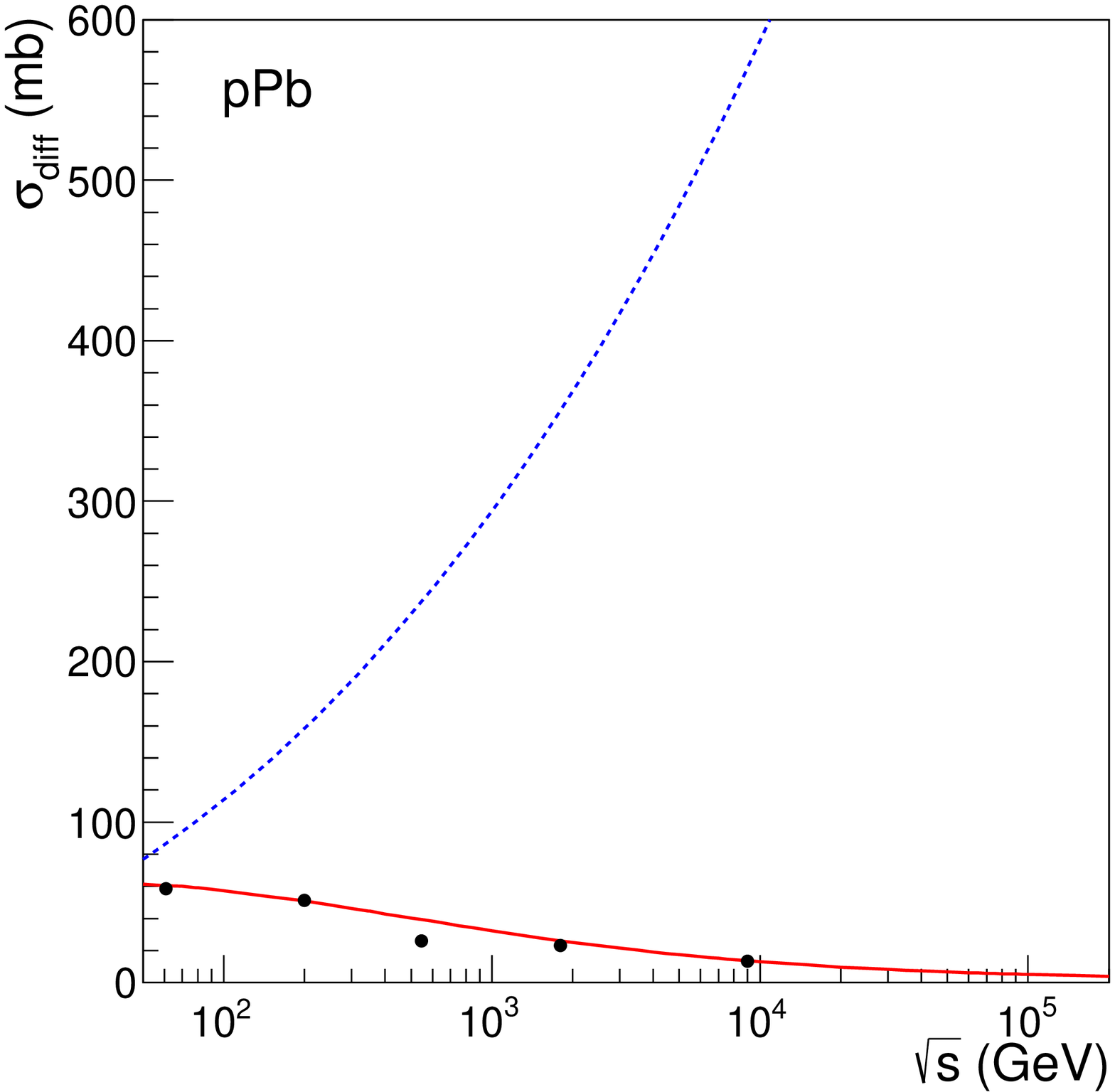}
 \includegraphics[width=0.31\textwidth]{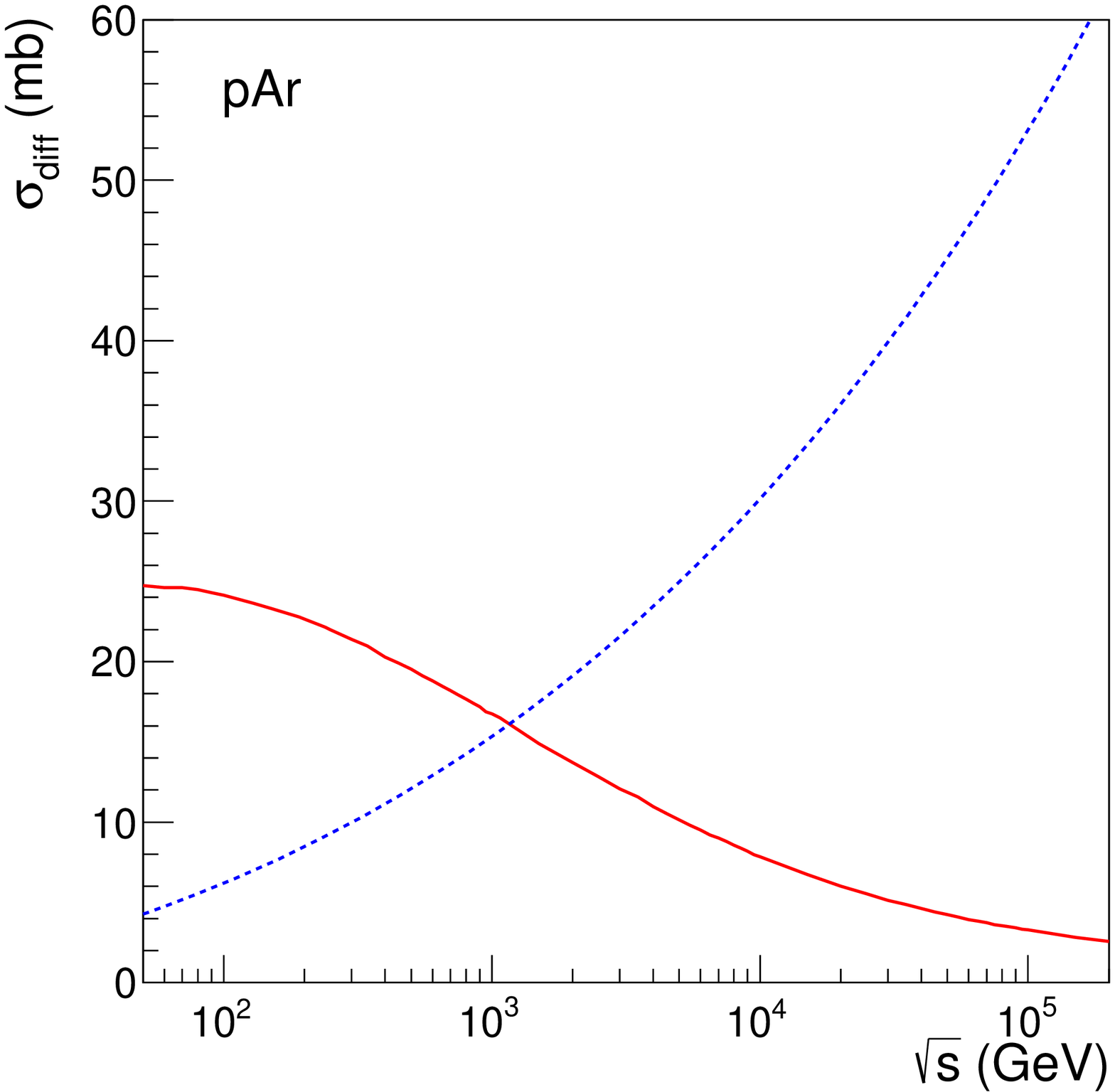}
  \includegraphics[width=0.31\textwidth]{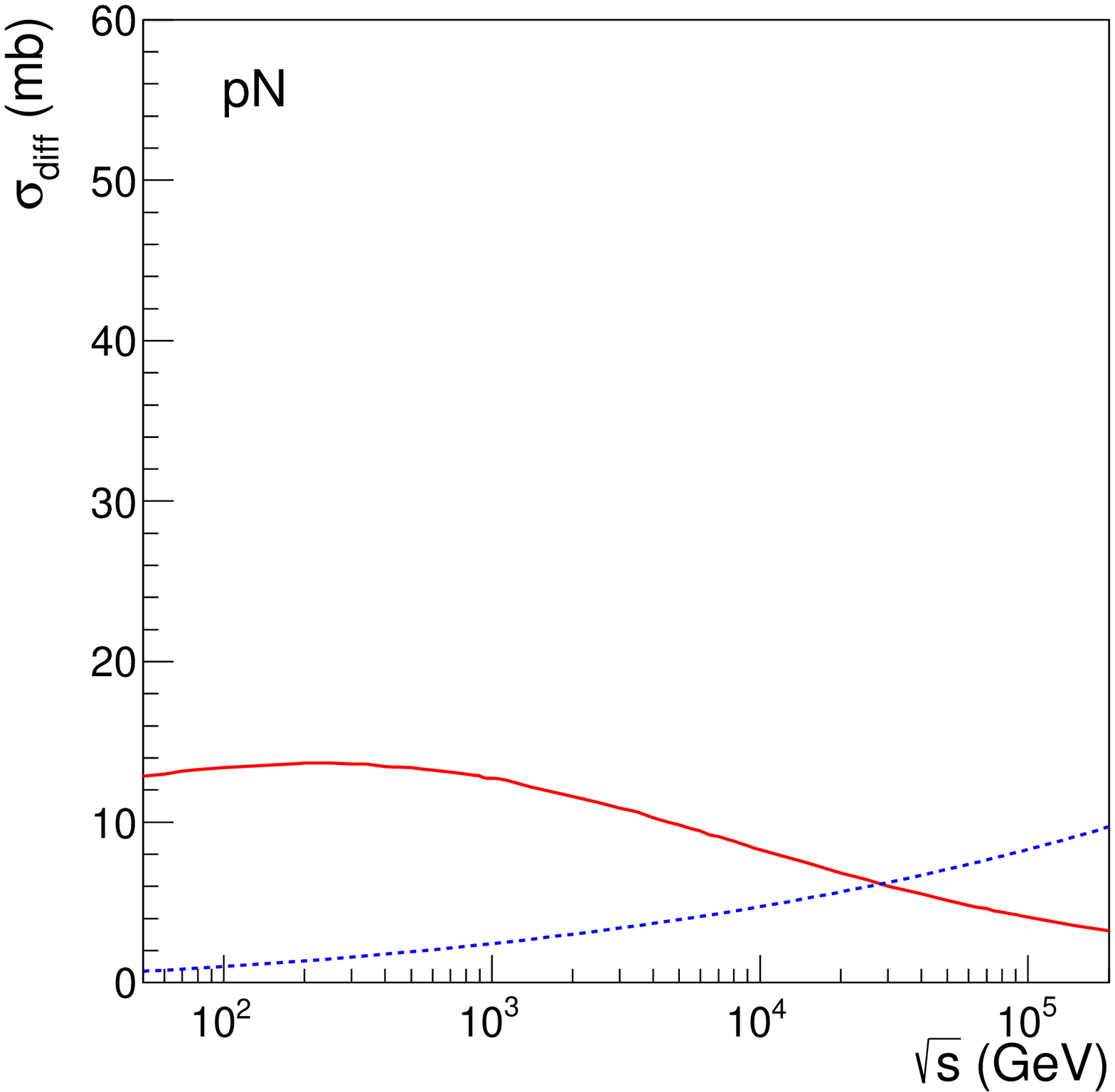}
 \caption{\label{fig:sigmadiff_pA} Diffractive dissociation cross section for
 proton -- lead (left), proton -- argon (central) and proton -- nitrogen (right) scattering 
 as a function of the energy in the center of mass system. 
 The cross sections due hadronic cross section fluctuations are represented by red solid curves
 and those from photon-induced interactions by blue dashed curves.
 In the $pPb$ case, the points represent the predictions for the diffractive excitation derived in Ref. ~\cite{Guzey:2006}.}
\end{figure}

Finally, we can estimate the magnitude of the diffractive excitation in $pA$ collisions at high energies. Our predictions are presented in  Fig.~\ref{fig:sigmadiff_pA} for proton -- lead, proton -- argon ($pAr$) and proton -- nitrogen scattering and are represented by the red solid lines. For comparison, we also present in the $pPb$ case, the predictions derived in Ref. ~\cite{Guzey:2006}. We have that the diffractive excitation cross section increases with the atomic nuclei and decreases with the energy, in agreement with the results derived in Ref. ~\cite{Guzey:2006}. In Ref. ~\cite{Guzey:2006} the authors have pointed out that the electromagnetic contribution for the diffractive excitation in $pA$ collisions becomes important. The basic idea is that in ultraperipheral collisions, the nucleus acts as a source of photons which interact with the proton \cite{upc}. This contribution can be estimated in terms of the nuclear photon flux ($n_A$) and the photon -- proton cross section ($\sigma_{\gamma p \rightarrow X}$) as follows:
\begin{eqnarray}
\sigma^{pA}_{e.m.} = \int \frac{d\omega}{\omega} n_A(\omega) \, \sigma_{\gamma p \rightarrow X} (\omega)\,\,,
\end{eqnarray}
where $\omega$ is photon energy. In our calculations we consider the same inputs used in  Ref. ~\cite{Guzey:2006} and the resulting predictions are represented in Fig. ~\ref{fig:sigmadiff_pA} by blue dashed curves. We have that the electromagnetic contribution increases with the energy and the atomic number, being dominant for $pPb$ collisions. For $pAr$ collisions, both contributions are similar at the LHC energies. Finally, it is important to emphasize that the electromagnetic contribution becomes dominant in $pN$ collisions at ultra high cosmic ray energies.

\section{Summary}
\label{sec:Sum}
  
The contribution of the diffractive excitation for the hadronic processes is directly related to the treatment of the internal degrees of freedom of the hadrons. During the last decades, several approaches have been proposed. In this paper we have considered the Good - Walker approach and updated the Miettinen -- Pumplin (MP) model considering the recent LHC data for the total and elastic $pp$ cross sections, as well older experimental data. We have demonstrated that this model is able to successfully describe the current data and a parametrization for the energy dependence of the main parameters of the MP model was proposed. The behavior of the cross sections for higher energies is analyzed and predictions for the energies of Run 3 of the LHC and those of the Cosmic Rays experiments have been derived. Our results demonstrated that the cross section for the diffraction excitation in $pp$ collisions is almost constant in the energy range probed by the LHC and slowly decreases at higher energies. Moreover, our results indicated that the Pumplin bound is not reached at the LHC and Cosmic Ray energies. We also have presented our results for the diffractive excitation in $pA$ collisions, as well have analyzed its impact on the predictions of the total and elastic cross sections considering different nuclei. The MP model, constrained by the $pp$ data, have been used to derive the main quantities present in the treatment of the diffractive excitation in $pA$ collisions. We demonstrated that the effect of fluctuations decreases at larger energies and heavier nuclei. Moreover, our results  indicated that the proton dissociation induced by photon interactions becomes dominant with the increasing of the energy and the atomic number.

\section*{Acknowledgements}
 This work was  partially financed by the Brazilian funding
agencies CNPq, CAPES,  FAPERGS and INCT-FNA (process number 
464898/2014-5).



\end{document}